# Random-mapped intensity optical neural network: all-optical two-layer computing for multimodal optical-field inference

Gi-Hyun Go[1], Doeon Lee[2], Gookho Song[1] and Mooseok Jang[1,*]

[1]Department of Bio and Brain Engineering, Korea Advanced Institute of Science and Technology (KAIST), Daejeon 34141, Republic of Korea

[2]School of Electrical Engineering, Hanyang University, Ansan 15588, Republic of Korea

Corresponding author: *mooseok@kaist.ac.kr

**Abstract**

Free-space optical neural networks offer distinct advantages for computational imaging and machine vision because they can compute directly on incident optical fields. However, conventional ONNs composed of cascaded linear optical components are bound to a general linear input–output relation with square-law detection between the input field $\mathbf{x}$ and output score $\mathbf{y}$, $\mathbf{y} = |\mathbf{Tx}|^2$, where at best every complex-valued element of the transmission matrix $\mathbf{T}$ is trainable. This restricts each output score to a quadratic form $\mathbf{x}^\dagger\mathbf{Ax}$ with rank-one decision matrix $\mathbf{A} = \mathbf{t}^\dagger\mathbf{t}$. Here, we present a random-mapped intensity optical neural network (RMI-ONN) as an all-optical two-layer computational network that lifts this rank-one limit. We numerically demonstrate that a high-dimensional feature projection by a disordered medium, a programmable nonnegative intensity mask, and segmented spatial power summation together can surpass the rank-one ceiling through the expressivity of higher-rank quadratic decision boundaries. Furthermore, exploiting the vectorial coherent wave-mixing nature of the disordered medium, we experimentally validate multimodal classification — amplitude, phase, and polarization — on MNIST, Fashion-MNIST, and Quick Draw with the RMI-ONN, under a single optical configuration across all encoding domains. These results provide a practical and conceptual basis for scalable direct-field optical processors capable of exploiting amplitude, phase, and polarization information within a unified intensity-based inference framework.

## Introduction

Optical computing has emerged as a compelling alternative for deep neural networks (DNNs), offering high bandwidth, massive parallelism, and fundamental reductions in processing latency by performing computations passively at the speed of light.[1–15] Depending on the application, optical neural network (ONN) architectures generally diverge into two primary paradigms: integrated chip-scale photonic circuits and free-space diffractive systems. Chip-scale approaches are particularly appealing for general-purpose computing and algorithmic acceleration.[15–17] Their compact footprint and compatibility with standard fabrication processes allow them to be tightly coupled with conventional electronic layers, forming powerful hybrid optoelectronic coprocessors. However, the required optical-to-electronic and electronic-to-optical conversions—utilizing modulators, photodetectors, and data converters—introduce significant latency, energy overhead, and coupling losses. Furthermore, forcing data through these planar integrated circuits could impose a strict dimensionality bottleneck.

By bypassing these structural and dimensionality bottlenecks, free-space architectures offer a distinct advantage for computational imaging and machine vision.[18–33] Because visual information in these tasks is inherently carried by incident optical wavefields, free-space ONNs can operate directly on high-dimensional spatial fields—performing computation in situ during propagation to entirely eliminate intermediate conversions. This direct-field processing is highly attractive for intelligent imaging front-ends, as it allows the network to seamlessly exploit wavelength, amplitude, phase, and polarization as complementary information channels.

Despite the inherent advantages of direct-field operation, the computational expressivity of many free-space ONNs based solely on passive linear propagation, including multi-layer diffractive neural networks ($D^2$NNs), remains constrained by the linearity of optical wave propagation. In the absence of nonlinear optical materials, the relationship between an input field $\mathbf{x}$ and the output field is strictly governed by a linear transmission matrix $\mathbf{T}$. Consequently, the system's sole nonlinearity is introduced only at the final sensor plane via square-law detection, yielding output intensities $\mathbf{y} = |\mathbf{Tx}|^2$. Mathematically, this restricts every output channel to a rank-one quadratic projection of the input field. This constraint severely limits the network's overall degrees of freedom, thereby preventing it from realizing the more expressive, higher-rank transformations needed to improve performance on demanding vision tasks. While existing approaches attempt to overcome this expressivity bottleneck by introducing physical optical nonlinearities (e.g., optical intensifiers, second-harmonic generation) or recurrent optoelectronic measurement-feedback loops,[33–38] these methods inevitably compromise the core benefits of free-space ONNs as they require immense power consumption, incur additional latency, or disrupt the ideal single-pass inference process.

Here, to overcome the fundamental expressivity bottleneck in linear optics, we propose a random-mapped intensity optical neural network (RMI-ONN)—a highly compact, all-optical framework that physically realizes a two-layer neural network using only a disordered scattering medium and an intensity spatial light modulator (SLM). In this scheme, the fixed disordered scattering medium naturally encodes incident amplitude, phase, and polarization information into a complex speckle intensity pattern. Consequently, the individual pixels comprising the spatial pattern on the SLM serve as the only learnable parameters, creating a reconfigurable higher-rank quadratic decision boundary. Numerical validation confirms this expressivity advantage, achieving 98% classification accuracy on MNIST and outperforming the corresponding single-projection baseline (i.e. $\mathbf{y} = |\mathbf{Tx}|^2$ where the elements of $\mathbf{T}$ are learnable), which yields approximately 93% under the same simulation setting. Within this framework, we experimentally demonstrate the classification of amplitude-, phase-, and polarization-encoded inputs across three benchmark datasets (MNIST, Fashion-MNIST, and Quick Draw) by reconfiguring only an intensity SLM, without involving any polarimetric or interferometric measurements. By merging passive high-dimensional optical scattering with programmable intensity modulation, this approach provides a scalable and efficient pathway toward reconfigurable direct-field optical processors for next-generation intelligent machine vision.

## Results

### Principle of Random-Mapped Intensity Optical Neural Network

Figure 1a summarizes the conventional free-space optical neural networks (ONNs) without nonlinear optical interactions. Regardless of how the linear optical system between the input fields and the sensor is configured, the relation between the input field and the output intensity can always be described by a transmission matrix followed by square-law detection: $\mathbf{y} = |\mathbf{T}_\mathrm{C}\mathbf{x}|^2$, where $\mathbf{x} \in \mathbb{C}^{N\times 1}$ is the input field, $\mathbf{y} \in \mathbb{R}^{C\times 1}$ is the output intensity, and $\mathbf{T}_\mathrm{C} \in \mathbb{C}^{C\times N}$ is the transmission matrix. For the $k$-th class, the resulting score is $y_k = |\mathbf{t}_k\mathbf{x}|^2 = \mathbf{x}^\dagger\mathbf{A}_k\mathbf{x}$, where $\mathbf{t}_k$ denotes the $k$-th row vector of $\mathbf{T}_\mathrm{C}$. Crucially, the corresponding decision matrix $\mathbf{A}_k = \mathbf{t}_k^\dagger\mathbf{t}_k$ is strictly rank-one, which fundamentally restricts the classifier's degrees of freedom and limits its MNIST accuracy to approximately 93%.

One can introduce a higher-dimensional nonlinear feature projection to overcome this restriction (Figure 1b). According to Cover's theorem on the separability of patterns, casting the data nonlinearly into a high-dimensional feature space dramatically increases the probability of the classes being linearly separable.[5,8,39] Optical random projections leverage this principle naturally: a disordered scattering medium combined with square-law detection executes a nonlinear feature lifting that can approximate high-dimensional kernel machines. In this framework, the relation between the input and the output vectors, $\mathbf{x}$ and $\mathbf{y}$, can be written as $\mathbf{y} = \mathbf{W}|\mathbf{T}_\mathrm{L}\mathbf{x}|^2$, where $\mathbf{T}_\mathrm{L} \in \mathbb{C}^{L\times N}$ is a fixed random transmission matrix and $\mathbf{W} \in \mathbb{R}^{C \times L}$ is a trainable digital readout matrix. Consequently, the decision matrix for the $k$-th class becomes a linear combination of random rank-one matrices:

$$\mathbf{B}_k = \sum_{l=1}^{L} (\mathbf{w}_k)_l \, \mathbf{t}_l^{\dagger} \mathbf{t}_l, \tag{1}$$

where $\mathbf{t}_l$ denotes the $l$-th row of $\mathbf{T}_\mathrm{L}$, and $(\mathbf{w}_k)_l$ is the weighting factor relating the $l$-th speckle feature to the $k$-th class score (i.e., the $l$-th element of the $k$-th row vector $\mathbf{w}_k$ of $\mathbf{W}$). This linear combination can increase the rank of the decision matrix $\mathbf{B}_k$ up to $L$. With a sufficiently large feature space, this rich basis set easily generates complex decision boundaries, achieving ~98% accuracy on MNIST. However, this architecture relies on a digital dense readout layer $\mathbf{W}$, so the class scores must still be computed by a trained digital matrix multiplication after measurement.

The RMI-ONN realizes a complete two-layer nonlinear network entirely in the optical domain by replacing this digital dense readout with a physically native intensity-domain operation, as shown in Figure 1c. At the readout stage, a spatial light modulator (SLM) applies a trainable intensity mask $\boldsymbol{\tau}$ to the high-dimensional features (i.e. speckle granules) projected by a disordered medium, implementing the per-mode non-negative weighting optically, and the weighted modes belonging to each class are directed onto a common detector. Crucially, because a photodetector accumulates photoelectrons in proportion to the optical intensity integrated over its active area, the detector performs sum pooling (equivalently, average pooling up to a constant) over the modes incident on it. The resulting operation is written as $\mathbf{y} = \mathbf{P}(\boldsymbol{\tau} \odot |\mathbf{T}_\mathrm{M}\mathbf{x}|^2)$, where $\mathbf{T}_\mathrm{M} \in \mathbb{C}^{M \times N}$ is a fixed random transmission matrix, $\odot$ denotes the element-wise product and $\mathbf{P} \in \{0,1\}^{C \times M}$ represents sum pooling operation for each class. To match the total number of trainable parameters between the dense digital readout in Figure 1b and the fully optical readout in Figure 1c, we set $M = LC$ and partition $\mathbf{T}_\mathrm{M}$ as $\mathbf{T}_\mathrm{M} = [\mathbf{T}^{(1)}; \mathbf{T}^{(2)}; \cdots \mathbf{T}^{(\mathrm{C})}]$, where $\mathbf{T}^{(k)} \in \mathbb{C}^{L \times N}$. The score for the $k$-th class is then written as the quadratic form $y_k = \mathbf{x}^{\dagger}\mathbf{Q}_k\mathbf{x}$, with the decision matrix

$$\mathbf{Q}_k = \sum_{l=1}^{L} (\boldsymbol{\tau}_k)_l \mathbf{t}_l^{(k)\dagger} \mathbf{t}_l^{(k)}, \tag{2}$$

where $\mathbf{t}_l^{(k)} \in \mathbb{C}^{1 \times N}$ denotes the $l$-th row of $\mathbf{T}^{(k)}$, and $(\boldsymbol{\tau}_k)_l$ denotes the $l$-th element of the class-wise trainable weights vector $\boldsymbol{\tau}_k \in \mathbb{R}_{\geq 0}^{L \times 1}$. Here, $\boldsymbol{\tau}_k$ is the corresponding subvector of $\boldsymbol{\tau} \in \mathbb{R}_{\geq 0}^{M \times 1}$, physically encoded on the SLM.

To benchmark this all-optical readout against its digital counterpart on equal footing, we first treat the intensity mask as unconstrained, $\boldsymbol{\tau} \in \mathbb{R}^{M \times 1}$, rather than $\boldsymbol{\tau} \in \mathbb{R}_{\geq 0}^{M \times 1}$. In this regime, the per-class decision matrices in Eqs. (1) and (2) are identical in construction: each is a real linear combination of $L$ rank-one Hermitian matrices generated from i.i.d. random complex-Gaussian vectors. Because $\mathbf{t}_l^{(k)}$ and $\mathbf{t}_l$ are drawn from the same distribution, the matrices $\mathbf{t}_l^{(k)\dagger}\mathbf{t}_l^{(k)}$ and $\mathbf{t}_l^\dagger\mathbf{t}_l$ span statistically equivalent subspaces of the $N \times N$ Hermitian matrix space, and an unconstrained $\boldsymbol{\tau}_k$ assumes exactly the role of the digital weights $\mathbf{w}_k$. The two networks thus parameterize the same family of quadratic decision functions, and their accuracies nearly coincide (Figure 1d). Notably, at $L = N \approx 784$ where the number of trainable weights coincides with that of the conventional rank-one quadratic classifier of Figure 1a, the higher-rank quadratic classifiers of Figures 1b and 1c already surpass the rank-one limit even though the trainable complex-valued $\mathbf{T}_\mathrm{C}$ carries twice as many ($2NC$) real degrees of freedom. As $L$ increases, the accuracy approaches a ~98% plateau in agreement with the well-known benchmark for low-degree polynomial-kernel support vector machines (SVMs) on raw MNIST pixels, consistent with the underlying degree-2 polynomial nature of the quadratic lifting.

To examine the effect of the physical constraint of non-negative intensity modulation, we compare the case of the unconstrained mask $\boldsymbol{\tau} \in \mathbb{R}^{M\times1}$ against the non-negative case $\boldsymbol{\tau} \in \mathbb{R}_{\geq0}^{M\times1}$. Considering that $\mathbf{Q}_k$ is the non-negative weighted combination for $\boldsymbol{\tau} \in \mathbb{R}_{\geq0}^{M\times1}$, $\mathbf{Q}_k$ is a positive semidefinite (PSD) matrix, confined to the PSD cone, a subset of the Hermitian space available to the unconstrained case. However, considering that the classification capability depends on the score differences for different classes, $y_k - y_j = \mathbf{x}^\dagger(\mathbf{Q}_k - \mathbf{Q}_j)\mathbf{x}$, and the random complex-Gaussian vectors $\mathbf{t}_l^{(k)}$ compose the rank-one PSD matrices with uniformly distributed directions, $\mathbf{Q}_k - \mathbf{Q}_j$ can be sampled across the full Hermitian space, including indefinite matrices, for both unconstrained and non-negative cases as $L$ grows. The non-negativity thus costs only representational efficiency, requiring more random features. As shown in Figure 1e, the constrained curve trails the unconstrained one at small $L$ and the two converge as the physically realizable RMI-ONN attains the full ~98% accuracy in the large-feature regime.

**Effective-rank analysis of decision matrix**

We next quantified the expressivity gained by increasing the feature dimension in the physically realizable RMI-ONN. Specifically, we computed the effective rank of the optimized decision matrices for MNIST, Fashion-MNIST, and Quick Draw, the three ten-class datasets used in the simulations (Figure 2a). Figure 2b shows the normalized eigenvalue spectra averaged over the ten class-wise decision matrices for each dataset, with error bars indicating the standard deviation across classes. When the feature dimension (i.e. the number of speckle) per class $L$ is small, the spectra are dominated by only a few large eigenvalues, indicating that the corresponding quadratic decision functions are governed by only a few dominant eigendirections. As $L$ increases, the eigenvalue spectrum broadens and decays more slowly, showing that the learned decision matrices become effectively higher-rank. This trend is consistently observed across all three datasets. We further quantified this spectral broadening using the effective rank $R_{\mathrm{eff}} = \exp(-\sum_i p_i \log p_i)$, where $p_i = |\lambda_i| / \sum_j |\lambda_j|$. This metric measures how broadly the eigenvalues are distributed.[40] As shown in Figure 2c, the classification accuracy increases with $R_{\mathrm{eff}}$, supporting the interpretation that broader quadratic representations improve classification performance. The improvement is rapid in the low-rank regime and becomes more gradual at high effective rank. Notably, the degree of saturation depends on the dataset: MNIST approaches saturation around $L = 784$, whereas the more complex Quick Draw dataset continues to benefit from further increases in effective rank.

The simulated confusion matrices in Figure 2d show the classification results at $L = 784$, yielding accuracies of 94.9% on MNIST, 85.3% on Fashion-MNIST, and 80.1% on Quick Draw. These results indicate that the proposed RMI-ONN can construct sufficiently expressive quadratic decision boundaries using only random optical mixing, square-law detection, and a nonnegative intensity readout. In particular, the effective-rank analysis supports the central mechanism of the architecture: increasing the number of random intensity features expands the accessible quadratic feature space and is associated with improved classification accuracy.

**Implementation of RMI-ONN**

Having established the expressivity of the RMI-ONN in simulations, we next implemented the physically realizable readout in an experimental free-space optical system. Figure 3a illustrates the optical implementation of the random-mapped intensity optical neural network (RMI-ONN). A 532-nm laser beam was used to generate the optical input field. Amplitude- and phase-encoded input fields were synthesized using a 4f relay system with a low-pass spatial filter.[41] The prepared input field was then scattered by a ground-glass diffuser, producing a high-dimensional speckle field at the plane of the transmission-type intensity SLM (Figure 3b). In our implementation, the feature dimension was fixed at $M = 784 \times 10 = 7840$, corresponding to $L = 784$ optical features assigned to each of the $C = 10$ class-specific output regions. To ensure wide random mixing and avoid localization of the effective mapping (i.e. non-diagonalized $\mathbf{T}$), the intensity SLM was placed 100 mm away from the diffuser, allowing light from each input mode to spread across the full modulation region. During inference, a trained nonnegative intensity mask, $\boldsymbol{\tau}$, was displayed on the transmission-type intensity SLM as an element-wise attenuation profile (Figure 3c). The intensity transmitted through the intensity SLM forms the weighted feature map $\boldsymbol{\tau} \odot |\mathbf{Tx}|^2$ (Figure 3d). This weighted intensity distribution is then optically integrated over predefined class-specific output regions, yielding the class scores $\mathbf{y} = \mathbf{P}(\boldsymbol{\tau} \odot |\mathbf{Tx}|^2)$ (Figure 3e). The predicted class is determined by the output region with the maximum summed intensity. In this proof-of-concept setup, an imaging sensor was used to record the full output plane for visualization and ROI-based intensity summation, which is equivalent to spatial power integration over predefined output regions. In a compact implementation, the same block-wise summation can be performed by an array of $C$ photodetectors, with one detector assigned to each class-specific output region.

## Validation of RMI-ONN on amplitude-encoded inputs

We first experimentally validated the RMI-ONN using amplitude-encoded inputs on three ten-class benchmark datasets: MNIST, Fashion-MNIST, and Quick Draw. To train the intensity-domain weights under real-valued nonnegative constraints, we adopted a hybrid calibration-training workflow (Figure 4a). Each $28 \times 28$ input amplitude image was synthesized by the SLM. The scattered intensity patterns were measured with the intensity SLM set to a uniform maximum-transmittance state, $\boldsymbol{\tau} = \mathbf{1}$, providing the experimentally acquired quadratic feature map $\boldsymbol{\phi} = |\mathbf{Tx}|^2$ with a vectorized input image $\mathbf{x} \in \mathbb{C}^{784\times1}$. The nonnegative readout mask $\boldsymbol{\tau}$ was then optimized in an equivalent digital model using the measured speckle features as fixed inputs and cross-entropy loss (see Methods for details). After training, the digitally optimized mask was uploaded to the intensity SLM, and inference was optically performed by the intensity measurement process where the individually weighted speckle patterns are summed in a block-wise manner. The optimization curves in Figure 4a show that the readout weights converge rapidly for all three datasets. The test accuracy monitored during the digital training process follows the training accuracy with a finite generalization gap, indicating stable optimization rather than overfitting-dominated behavior. This gap is small for MNIST and Fashion-MNIST, whereas it is more pronounced for Quick Draw, consistent with the larger intra-class variability and more complex decision boundaries of sketch-based data.

The confusion matrices in Figure 4b summarize the experimental classification results for amplitude-encoded inputs. The RMI-ONN achieves accuracies of 85.0%, 78.1%, and 65.2% for MNIST, Fashion-MNIST, and Quick Draw, respectively. These results demonstrate that a fixed disordered medium, followed by a physically implemented nonnegative intensity readout, can perform image classification using directly encoded amplitude information without requiring a complex electronic readout layer.

## Validation of RMI-ONN on phase- and polarization-encoded inputs

We next examined whether the same RMI-ONN architecture can process optical information encoded in the phase and polarization domains, which typically require dedicated optical modulation techniques in conventional ONNs. The scattering medium inherently multiplexes this multidimensional field information into naturally accessible high-dimensional speckle features. For phase encoding, each $28 \times 28$ grayscale image was encoded as a two-dimensional spatial phase pattern, as shown in Figure 5a. In this encoding, the grayscale values were mapped to phase delays ranging from 0 to $\pi$. In vectorized form, the phase-encoded input field is written as $\mathbf{x} = \exp(\mathrm{i}\pi\mathbf{G})$, where $\mathbf{G} \in [0,1]^{784\times1}$ denotes the vectorized grayscale image. In this configuration, the incident amplitude was kept approximately uniform, and the class-discriminative information was carried primarily by the spatial phase distribution. After coherent random mixing by the disordered medium, phase variations were converted into measurable intensity variations through square-law detection, yielding the quadratic intensity feature map $\boldsymbol{\phi} = |\mathbf{Tx}|^2$. The $m$-th speckle feature can be expanded as $\phi_m = |(\mathbf{Tx})_m|^2 = \sum_{s,t} T_{ms} T_{mt}^* x_s x_t^*$, showing that the detected intensity contains second-order interaction terms between input field components. For phase-encoded inputs, these terms depend on relative phase differences between input modes, allowing phase-dependent information to be transferred into a measurable speckle intensity representation without interferometric field reconstruction. The test-set monitoring curves in Figure 5b show stable convergence of the nonnegative readout for phase-encoded inputs across MNIST, Fashion-MNIST, and Quick Draw. The corresponding confusion matrices in Figure 5c yield classification accuracies of 86.5%, 80.5%, and 66.9%, respectively. The comparable performance between amplitude and phase encodings supports the interpretation that coherent random mixing and square-law photodetection embed phase-dependent information into the measured intensity feature space. Thus, the RMI-ONN can exploit phase-encoded information using the same physically implemented intensity-domain readout.

To further extend the framework beyond scalar-field encodings, we performed classification using polarization-encoded inputs, where class-discriminative information is carried by spatially varying polarization states. In this case, the input field is represented by two orthogonal polarization components, $\mathbf{x} = \begin{bmatrix} \mathbf{x}_\mathrm{H} \\ \mathbf{x}_\mathrm{V} \end{bmatrix} \in \mathbb{C}^{2N}$, where $\mathbf{x}_\mathrm{H}, \mathbf{x}_\mathrm{V} \in \mathbb{C}^{N\times 1}$ are the complex field coefficients in the H- and V-polarization bases, respectively. In the most general form, propagation through the scattering medium can be described by a polarization-resolved transmission matrix, $\mathbf{T}_\mathrm{pol} = \begin{bmatrix} \mathbf{T}_\mathrm{HH} & \mathbf{T}_\mathrm{HV} \\ \mathbf{T}_\mathrm{VH} & \mathbf{T}_\mathrm{VV} \end{bmatrix} \in \mathbb{C}^{2M\times 2N}$, where each sub-block $\mathbf{T}_{pq} \in \mathbb{C}^{M\times N}$ describes coupling from input polarization $q$ to output polarization $p$. In our implementation, the entrance polarizer of the transmission-type intensity SLM acts as a fixed linear analyzer before intensity modulation and detection. With the analyzer aligned to the H basis, the polarization-resolved scattering process is reduced to the analyzer-projected effective transmission matrix $\mathbf{T}_\mathrm{eff} = \mathbf{a}_\mathrm{H} \mathbf{T}_\mathrm{pol} = [\mathbf{T}_\mathrm{HH} \quad \mathbf{T}_\mathrm{HV}] \in \mathbb{C}^{M\times 2N}$. The measured feature map therefore retains the same quadratic form as in the scalar-field case, $\boldsymbol{\phi} = |\mathbf{T}_\mathrm{eff}\mathbf{x}|^2$. Because $\mathbf{T}_\mathrm{eff}$ contains contributions from both input polarization components, the resulting quadratic intensity features can include polarization-dependent interactions between the H and V components within the analyzer-selected output channel. This allows vectorial field information to be embedded into the measured intensity features while using the same nonnegative intensity-domain readout as in the amplitude- and phase-encoded cases.

As shown in Figure 5d, spatially varying polarization inputs were generated by mapping each grayscale value to a local linear-polarization angle ranging from 0 to $\pi/2$. Experimentally, these input fields were synthesized using a nematic liquid crystal array combined with a linear polarizer, producing spatially varying polarization states across the input plane. In vectorized form, the polarization-encoded input field is represented as $\mathbf{x} = \begin{bmatrix} \cos(\pi\mathbf{G}/2) \\ \sin(\pi\mathbf{G}/2) \end{bmatrix}$. The test-set monitoring curves in Figure 5e show stable convergence of the nonnegative readout weights for polarization-encoded inputs. The corresponding confusion matrices in Figure 5f show classification accuracies of 81.6%, 72.1%, and 59.1% for MNIST, Fashion-MNIST, and Quick Draw, respectively.

To benchmark the experimental performance against an idealized random-feature model, we also performed numerical simulations in which the transformation matrix for each output basis was modelled as a random Gaussian matrix. Across all phase- and polarization-encoded tasks, the simulated and experimental accuracies differed by around 10%. The observed discrepancy between simulation and experiment can be attributed to practical nonidealities in the optical setup, including measurement noise, residual calibration errors during intensity modulation, and mechanical drift between the training and inference phases. Nevertheless, the successful classification of phase- and polarization-encoded inputs confirms that the same scattering-based front-end and physically implemented intensity-domain readout can process amplitude, phase, and polarization information within a unified quadratic intensity-mapping framework.

**Discussion**

We have introduced a random-mapped intensity optical neural network (RMI-ONN) that implements a two-layer neural network with a quadratic activation function in an all-optical manner. Unlike a single-layer counterpart represented in $\mathbf{y} = |\mathbf{Tx}|^2$, such as in D2NN, where each output channel is restricted to a rank-one quadratic decision matrix, the RMI-ONN introduces a trainable intensity-domain readout $\boldsymbol{\tau}$ that manipulates higher-rank quadratic decision boundaries based on the input-to-score relation $\mathbf{y} = \mathbf{P}(\boldsymbol{\tau} \odot |\mathbf{Tx}|^2)$. Remarkably, the multimodal classification capability of the RMI-ONN was experimentally validated on amplitude-, phase-, and polarization domains without explicit interferometric reconstruction or polarization-resolved detection. This capability arises because the speckle features generated through coherent scattering and mutual interference process inherently retain information about amplitude correlations, relative phase differences, and inter-polarization correlations. These results show that the RMI-ONN is not merely an intensity-image classifier, but a unified direct-field classifier for multiple optical degrees of freedom.

The presented architecture highlights the usefulness of disordered photonics as a compact physical fan-out layer, which is conventionally implemented using a micro-lens array. A single scattering medium distributes information from each input mode across many output speckle modes, creating a redundant and element-wise addressable intensity representation. The trainable mask $\boldsymbol{\tau}$ then selects and weights these features, while the class-wise summation operation (i.e. areal accumulation of photoelectrons) converts spatially distributed optical power into decision scores. In the present proof-of-concept system, a camera is used to record and analyze the output plane, but the same block-wise aggregation could in principle be implemented using a small number of photodetectors, suggesting a path toward compact optical front-ends in which high-dimensional optical fields are processed before electronic digitization.

More broadly, our implementation of a two-layer optical computing network can be regarded as a physical readout primitive for intensity-domain optical neural networks. When combined with a programmable optical transformation, such as a trainable diffractive layer, metasurface, or spatial light modulator, the same readout concept could support architectures in which both the coherent optical feature generator and the intensity-domain readout are trainable. In such a system, the forward model would move from a fixed random mapping $\mathbf{y} = \mathbf{P}(\boldsymbol{\tau} \odot |\mathbf{Tx}|^2)$ toward a more flexible form $\mathbf{y} = \mathbf{P}(\boldsymbol{\tau} \odot |\mathbf{T}_\theta\mathbf{x}|^2)$, where $\mathbf{T}_\theta$ denotes a trainable optical transformation. With additional fan-out, multi-head readouts, or differential detection schemes for signed weighting, this readout primitive could serve as a general intensity-domain interface for scalable optical computing platforms, including free-space optical processors, metasurface-based systems, and chip-scale photonic neural networks.[42]

Several limitations remain. First, the current system requires a hybrid calibration-training procedure: speckle features are experimentally measured with a uniform mask, the nonnegative readout weights are optimized digitally, and the optimized mask is then uploaded to the intensity SLM for optical inference. This workflow directly implements the trained readout in hardware, but it still relies on digital optimization and assumes stability of the measured feature map. Second, the performance of speckle-based processors depends on mechanical and thermal stability, because drift of the scattering medium or optical alignment changes the effective mapping $\mathbf{T}$. Finally, multiple scattering and intensity modulation introduce optical loss and finite dynamic range, which can reduce the signal-to-noise ratio and enlarge the gap between idealized simulations and experimental measurements.

Overall, the RMI-ONN demonstrates that the disordered mixing process in coherent domain and the trainable modulation in incoherent domain can be seamlessly combined into a physically implementable quadratic feature-mapping architecture for all-optical inference. This provides a practical and conceptual basis for future direct-field optical processors capable of exploiting amplitude, phase, and polarization information within a unified intensity-based inference framework.

## Materials and methods

**Implementation of the random-mapped intensity optical neural network**

The random-mapped intensity optical neural network (RMI-ONN) was implemented by aligning and calibrating a transmission-type intensity spatial light modulator (SLM; resolution: 1280×800 pixels; pixel size: ~10 μm) that includes an entrance linear polarizer which projects the scattered vector field onto a fixed analyzer state (aligned to H in our basis), yielding the effective mapping $\mathbf{T}_{\mathrm{eff}} = \mathbf{a}_{\mathrm{H}}\mathbf{T}_{\mathrm{pol}}$. To increase the output intensity per channel, each output mode was assigned to a 2×2 group of SLM pixels instead of a single pixel. To achieve pixel matching between the SLM and the imaging sensor (BFS-U3-50S5M-C, FLIR; resolution: 2448×2048 pixels; pixel size: 3.45 μm), the magnification of the imaging system was adjusted so that one SLM pixel corresponded to three camera pixels. An iris placed between the lenses of the 4f imaging system was used to set the speckle size larger than that of a single output channel. Calibration of the intensity SLM was performed to confirm the correct operation of each channel. To evaluate potential cross-talk between neighboring outputs, the SLM plane was divided into odd and even regions, and calibration was carried out separately for each region. The results confirmed that cross-talk between adjacent outputs was negligible. After digital training, the optimized nonnegative intensity mask was uploaded to the intensity SLM, and the corresponding optical intensities were experimentally measured.

**Training and evaluation of the RMI-ONN**

The trainable intensity-domain weights of the RMI-ONN were optimized digitally using experimentally measured speckle intensity features and were subsequently implemented optically on the intensity SLM. For each dataset and encoding condition, the optical feature maps were first measured with the intensity SLM set to a uniform maximum-transmittance state, $\boldsymbol{\tau} = \mathbf{1}$. For Fashion-MNIST and Quick Draw, the measured datasets consisted of 60,000 training samples and 10,000 test samples. For MNIST, class-balanced subsets containing 54,210 training samples and 8,920 test samples were used, corresponding to 5,421 training samples and 892 test samples per class. This balancing was performed because the original MNIST dataset contains different numbers of samples for different digit classes. To reduce possible temporal bias during optical measurement, the samples were not acquired class by class. Instead, they were measured in an interleaved cyclic order across classes. This acquisition protocol avoided associating a specific class with a particular measurement time, thereby reducing the possibility that slow temporal drift in the optical system could be learned as a class-dependent feature.

For the $n$-th input field $\mathbf{x}^{(n)}$, the measured speckle feature vector is expressed as $\boldsymbol{\phi}^{(n)} = \left|\mathbf{T}\mathbf{x}^{(n)}\right|^2 \in \mathbb{R}_{\geq 0}^{M}$, where $\mathbf{T}$ denotes the experimentally realized optical transformation from the input field to the speckle plane. In the implemented system, the speckle features were arranged into ($C = 10$) class-specific output regions, each containing ($L = 784$) intensity features. Therefore, the total number of measured features was $M = L \times C = 784 \times 10 = 7840$. The RMI-ONN output for the n-th sample is given by $\mathbf{y}^{(n)} = \mathbf{P}(\boldsymbol{\tau} \odot \boldsymbol{\phi}^{(n)})$, where $\boldsymbol{\tau}$ is the trainable nonnegative intensity mask, $\odot$ denotes element-wise multiplication, and $\mathbf{P}$ denotes class-wise block summation. In this operation, each output score is obtained by summing the weighted speckle intensities assigned to the corresponding class-specific output region.

The measured speckle intensity patterns were used as fixed input features in an equivalent digital model. The optical transformation $\mathbf{T}$ and the measured features $\boldsymbol{\phi}^{(n)}$ were not updated during training; only the intensity mask $\boldsymbol{\tau}$ was optimized under a real-valued nonnegative constraint. Before training, the measured grayscale feature images were normalized by the camera intensity range and reshaped into 7840-dimensional feature vectors. The class-wise output scores $\mathbf{y}^{(n)} = \mathbf{P}(\boldsymbol{\tau} \odot \boldsymbol{\phi}^{(n)})$ were treated as logits, and $\boldsymbol{\tau}$ was optimized by minimizing the sparse categorical cross-entropy loss over the measured training set: $\mathcal{L}(\boldsymbol{\tau}) = \frac{1}{N_{\mathrm{train}}} \sum_{n=1}^{N_{\mathrm{train}}} \left[ \log\left( \sum_{k=1}^{C} \exp\left( y_k^{(n)} \right) \right) - y_{c^{(n)}}^{(n)} \right]$. Here, $N_{\mathrm{train}}$ denotes the number of measured training samples and $c^{(n)}$ denotes the ground-truth label of the $n$-th training sample. The optimization was performed using the Adam optimizer for 50 epochs. The measured training set was used to update the trainable mask $\boldsymbol{\tau}$, whereas the measured test set was held out from optimization and used only for performance evaluation.

After optimization, the learned nonnegative mask $\boldsymbol{\tau}$ was uploaded to the intensity SLM. The final RMI-ONN inference was performed by applying the optimized intensity-domain mask to the speckle intensity features and summing the weighted optical power over the class-specific output regions. The predicted label was determined by selecting the output region with the maximum summed intensity, $\hat{c}^{(n)} = \underset{k}{\operatorname{argmax}}\, y_k^{(n)}$. The final classification accuracy and confusion matrix were calculated using the corresponding measured test set. This hybrid calibration-training procedure ensures that the trainable readout mask is optimized from experimentally measured optical feature maps while remaining directly implementable in the physical RMI-ONN hardware through nonnegative intensity modulation and class-wise spatial power summation.

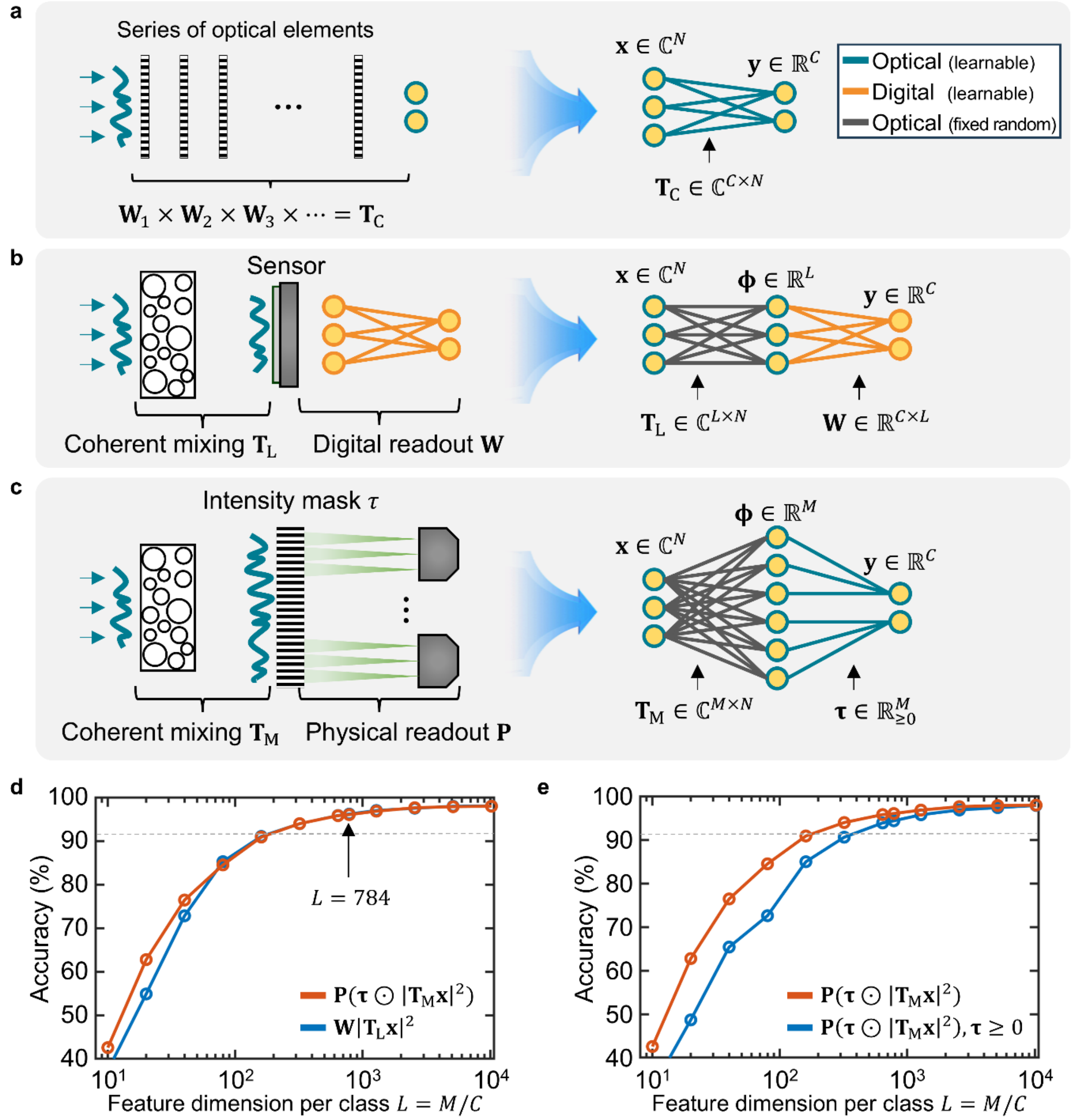

**Figure 1. Quadratic intensity-mapping framework for direct optical-field ONNs. a** Conventional free-space coherent optical classifier, where an effective transmission matrix $\mathbf{T}_\mathrm{C}$ and square-law detection produce class scores $\mathbf{y} = |\mathbf{T}_\mathrm{C}\mathbf{x}|^2$, with each score corresponding to a rank-one quadratic decision matrix. **b** Random-feature optical kernel model, where a fixed scattering-induced mapping generates high-dimensional quadratic intensity features $|\mathbf{T}_\mathrm{L}\mathbf{x}|^2$, followed by a dense digital readout $\mathbf{W}$ that enables higher-rank quadratic decision matrices. **c** Proposed RMI-ONN, where the dense digital readout is replaced by element-wise intensity modulation $\boldsymbol{\tau}$ and block-wise incoherent spatial summation $\mathbf{P}$, yielding $\mathbf{y} = \mathbf{P}(\boldsymbol{\tau} \odot |\mathbf{T}_\mathrm{M}\mathbf{x}|^2)$. This physically native readout realizes higher-rank quadratic decision matrices using nonnegative intensity weights and spatial summation. **d** Simulated MNIST accuracy as a function of the per-class feature dimension $L$, comparing the dense random-feature readout and the structured RMI-ONN readout under unconstrained real-valued weights. **e** Effect of the nonnegative intensity constraint on the RMI-ONN readout. In both panels, the gray dotted line indicates the accuracy of the single-projection baseline model, $\mathbf{y} = |\mathbf{T}_\mathrm{C}\mathbf{x}|^2$.

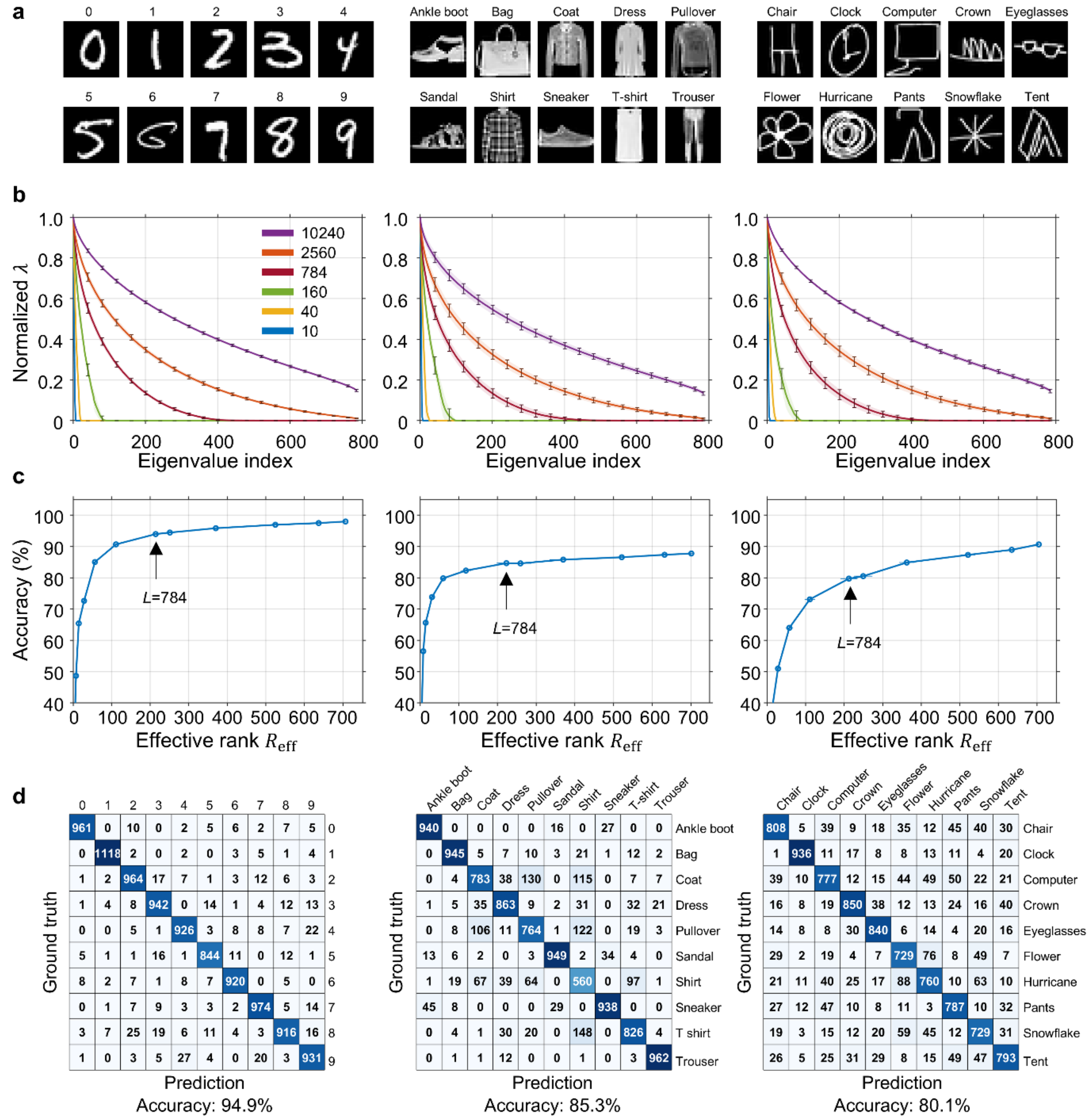

**Figure 2. Effective-rank analysis of the random-mapped intensity optical neural network. a** Representative examples from the three ten-class datasets used in the simulations: MNIST, Fashion-MNIST, and Quick Draw. **b** Normalized eigenvalue spectra of the learned decision matrices for different numbers of random features per class, $L$. The spectra are averaged over ten classes, and the error bars indicate the standard deviation. As $L$ increases, the eigenvalue distribution becomes broader, indicating that the learned decision matrices become effectively higher-rank. **c** Classification accuracy as a function of the effective rank $R_{\mathrm{eff}}$. The arrows indicate $L = 784$, where the number of random features per class is comparable to the input dimension. Higher $R_{\mathrm{eff}}$ is associated with improved accuracy, and the performance gradually saturates at high effective rank. **d** Simulated confusion matrices obtained at $L = 784$ for MNIST, Fashion-MNIST, and Quick Draw, showing classification accuracies of 94.9%, 85.3%, and 80.1%, respectively.

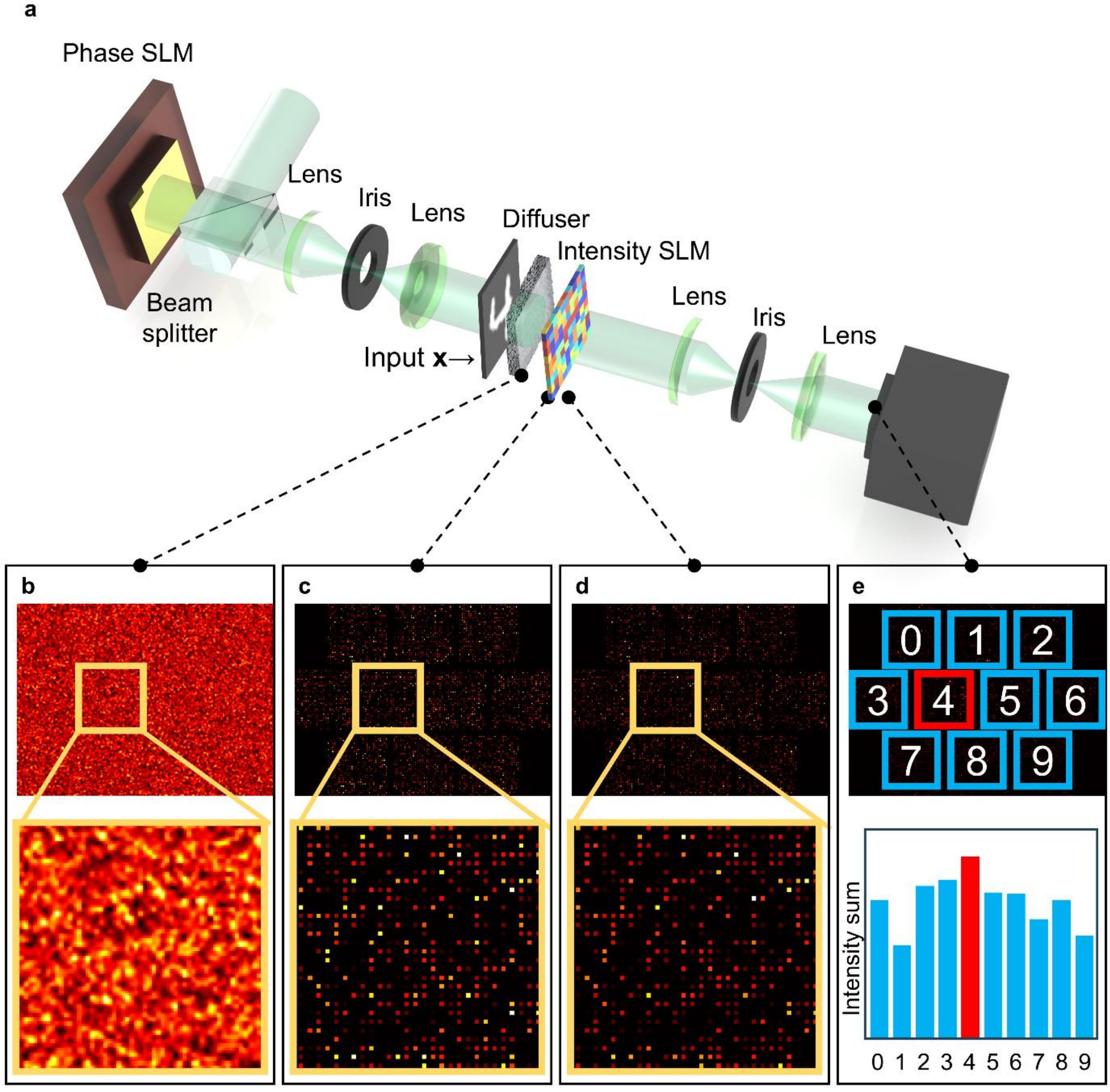

**Figure 3. Experimental implementation and inference pathway of the RMI-ONN. a** Optical setup. A 532-nm laser beam is shaped by a phase SLM to generate an input field $\mathbf{x}$ for amplitude- or phase-encoded operation, scattered by a ground-glass diffuser, modulated by a transmission-type intensity SLM, and recorded by an imaging sensor. **b** Representative speckle intensity feature map, $|\mathbf{Tx}|^2$, measured with the intensity SLM set to a uniform maximum-transmittance state, $\boldsymbol{\tau} = \mathbf{1}$. **c** Trained nonnegative intensity mask, $\boldsymbol{\tau}$, displayed on the intensity SLM as an element-wise attenuation profile. **d** Weighted intensity feature map, $\boldsymbol{\tau} \odot |\mathbf{Tx}|^2$, formed after transmission through the intensity SLM. **e** Block-wise incoherent summation over $C = 10$ predefined output regions yields class scores $\mathbf{y} = \mathbf{P}(\boldsymbol{\tau} \odot |\mathbf{Tx}|^2)$; the predicted label corresponds to the output region with the maximum summed intensity.

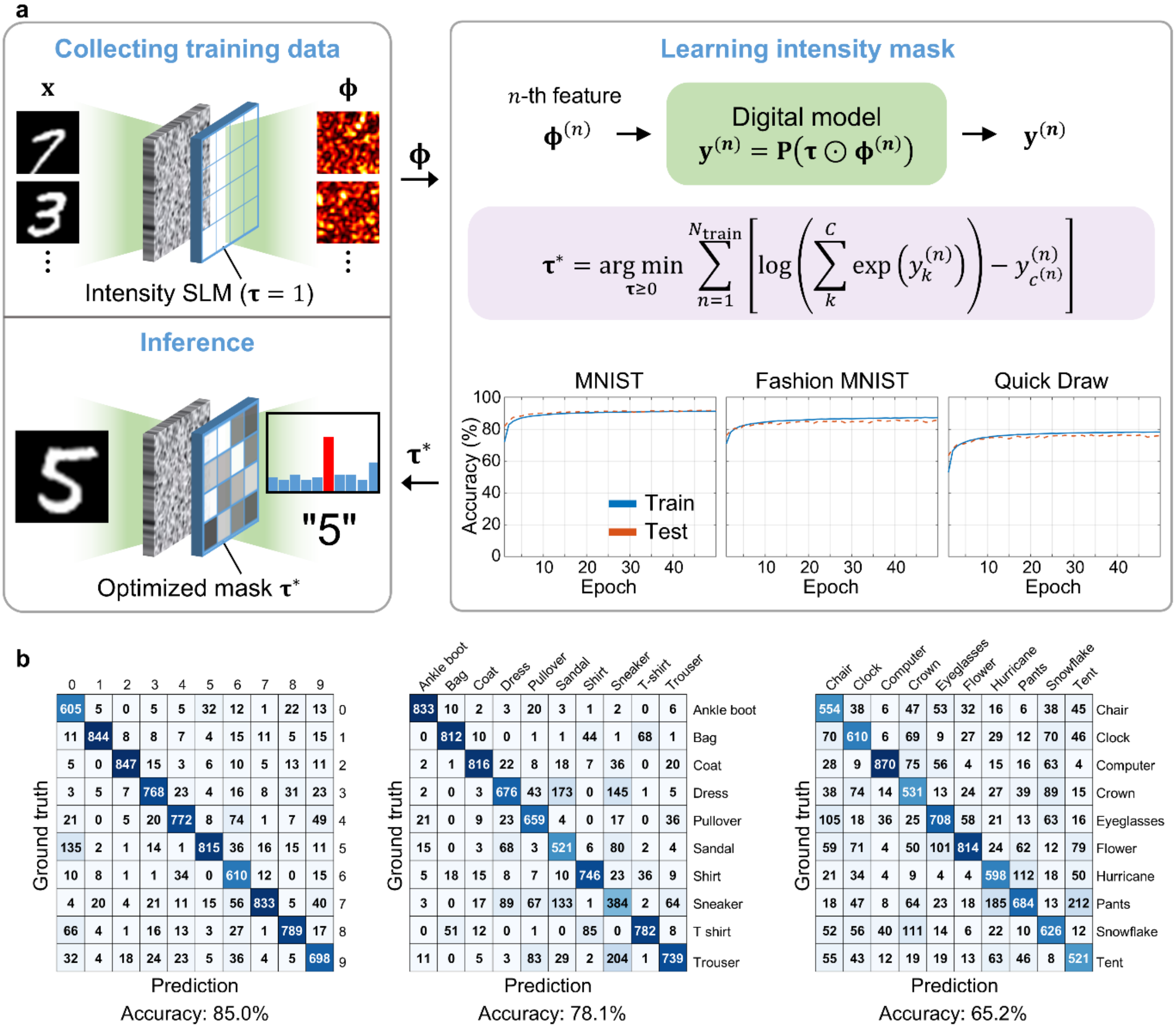

**Figure 4. Experimental validation of the RMI-ONN using amplitude-encoded inputs. a** Hybrid calibration-training workflow for amplitude-encoded classification. Each input image is encoded as an amplitude input field $\mathbf{x}$. With the intensity SLM set to a uniform maximum-transmittance state, $\boldsymbol{\tau} = \mathbf{1}$, the scattered intensity features $\boldsymbol{\phi} = |\mathbf{Tx}|^2$ are experimentally measured and used as fixed training features to optimize the nonnegative readout mask $\boldsymbol{\tau}$ in an equivalent digital model. The optimized mask $\boldsymbol{\tau}^*$ is subsequently uploaded to the intensity SLM for optical inference. The training curves show stable convergence for MNIST, Fashion-MNIST, and Quick Draw. **b** Confusion matrices for amplitude-encoded inputs. The RMI-ONN achieves classification accuracies of 85.0%, 78.1%, and 65.2% on MNIST, Fashion-MNIST, and Quick Draw, respectively.

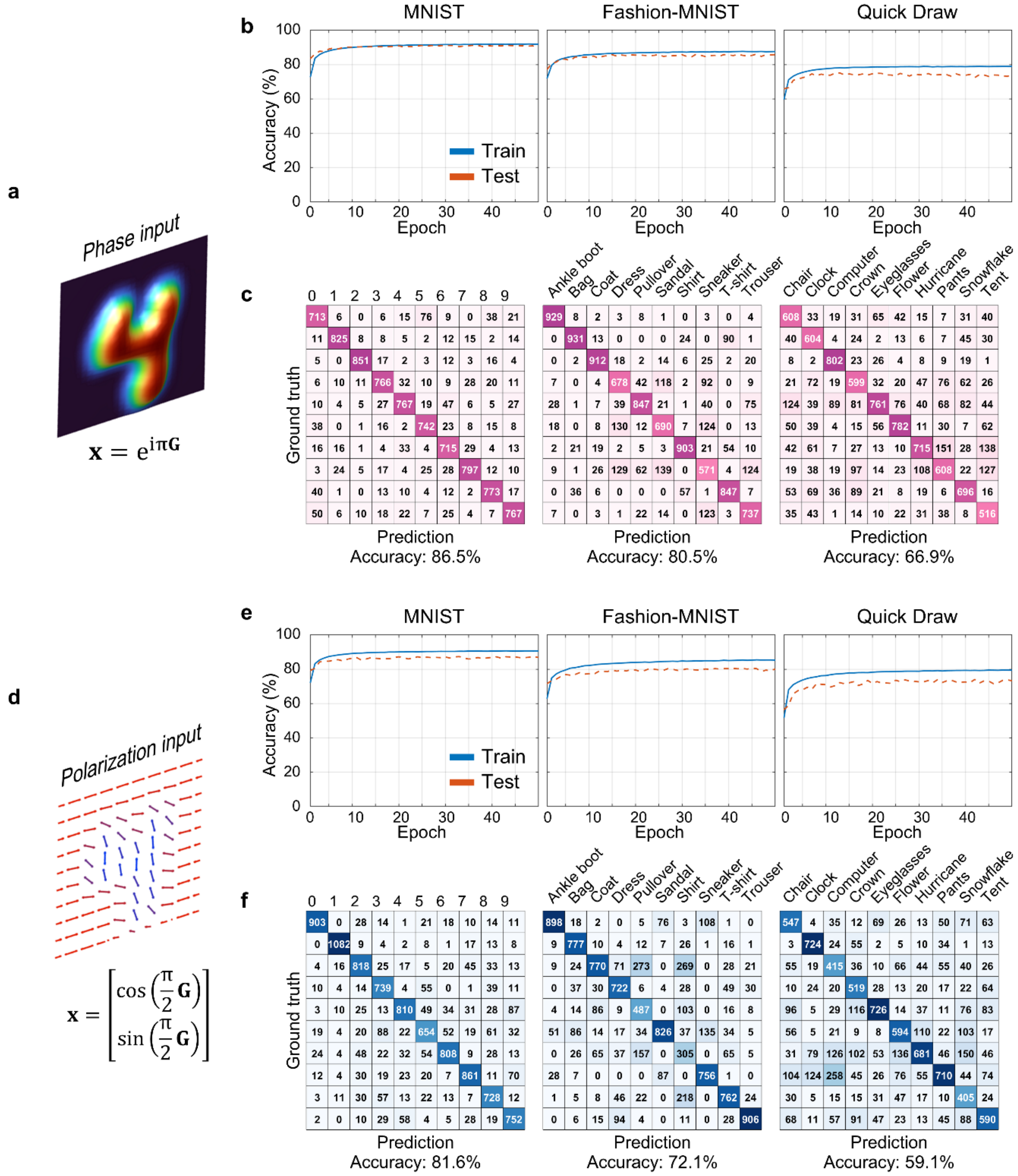

**Figure 5. Experimental validation of the RMI-ONN using phase- and polarization-encoded inputs. a** Each 28 × 28 grayscale image is encoded as a two-dimensional spatial phase pattern, with grayscale values mapped to phase delays ranging from 0 to π. In vectorized form, the phase-encoded input field is written as $\mathbf{x} = \exp(i\pi\mathbf{G})$, where $\mathbf{G} \in [0,1]^{784\times1}$ denotes the vectorized grayscale image. **b** Test-set monitoring curves for phase-encoded classification on MNIST, Fashion-MNIST, and Quick Draw. **c** Confusion matrices for phase-encoded inputs, showing classification accuracies of 86.5%, 80.5%, and 66.9% on MNIST, Fashion-MNIST, and Quick Draw, respectively. **d** Spatially varying polarization inputs are generated by mapping each grayscale value to a local linear-polarization angle ranging from 0 to π/2. In vectorized form, the polarization-encoded input field is represented as $\mathbf{x} = \begin{bmatrix} \cos(\pi\mathbf{G}/2) \\ \sin(\pi\mathbf{G}/2) \end{bmatrix}$. **e** Test-set monitoring curves for polarization-encoded classification on MNIST, Fashion-MNIST, and Quick Draw. **f** Confusion matrices for polarization-encoded inputs, showing classification accuracies of 81.6%, 72.1%, and 59.1%, respectively. These results demonstrate that the same scattering-based front-end and intensity-domain readout can process vectorial optical-field information without polarization-resolved detection or interferometric field reconstruction.